\documentclass[12pt]{article}
\usepackage{axodraw}

\def\ltap{\ \raisebox{-.4ex}{\rlap{$\sim$}} \raisebox{.4ex}{$<$}\ }
\def\gtap{\ \raisebox{-.4ex}{\rlap{$\sim$}} \raisebox{.4ex}{$>$}\ }

\begin{document}

\begin{titlepage}
\noindent
\begin{minipage}[t]{6in} 
\begin{flushright}
 KEK-TH-726        \\
 SISSA 110/2000/EP \\
 UT-918            \\
 UCB-PTH-00/39     \\
 LBNL-47173        \\
 hep-ph/0012118    \\
\end{flushright} 
\end{minipage}

\begin{center}
{\Large
{\bf  Variations on Supersymmetry Breaking}                       \\
{\bf   and Neutrino Spectra}}                                     \\[4ex]
{\bf F.~Borzumati$^{\,a,b}$, K.~Hamaguchi$^{\,c}$, 
     Y.~Nomura$^{\,d,e}$, and T.~Yanagida$^{\,c,f}$}              \\[2ex]
$^a$ {\it Theory Group, KEK, Tsukuba, Ibaraki 305-0801, Japan}    \\
$^b$ {\it Scuola Internazionale Superiore di Studi Avanzati 
      (SISSA), Via Beirut 4, I-34014 Trieste, Italy}              \\
$^c$ {\it Department of Physics, University of Tokyo,
          Tokyo 113-0033, Japan}                                  \\
$^d$ {\it Department of Physics, University of California, Berkelely, 
          California 94720, USA}                                  \\
$^e$ {\it Theoretical Physics Group, Lawrence Berkeley National 
          Laboratory, Berkelely, California 94720, USA}           \\
$^f$ {\it Research Center for the Early Universe,
          University of Tokyo, Tokyo 113-0033, Japan}             \\[10ex]
\end{center}
{\begin{center} ABSTRACT \end{center}}
\vspace*{1mm}

\parbox{14.5cm}
{
\noindent
The problem of generating light neutrinos within 
supersymmetric models is discussed. It is shown that 
the hierarchy of scales induced by supersymmetry breaking 
can give rise to suppression factors of the correct 
order of magnitude to produce experimentally allowed
neutrino spectra. 
}
\vfill

{\begin{center} 
{\it Talk presented by F. Borzumati at the } \\
{\it Post Summer Institute Workshop on ``Neutrino Physics''}\\
{\it  Fuji-Yoshida, Japan, August 21-24, 2000}
\end{center}}

\end{titlepage}
\thispagestyle{empty}

\newpage
\phantom{a}
\vfill 
\thispagestyle{empty}

\newpage 
\setcounter{page}{1}


\begin{center}
{\Large
{\bf  Variations on Supersymmetry Breaking}                       \\
{\bf   and Neutrino Spectra}}                                     \\[3ex]
{\bf F.~Borzumati$^{\,a,b}$, K.~Hamaguchi$^{\,c}$, 
     Y.~Nomura$^{\,d,e}$, and T.~Yanagida$^{\,c,f}$}                \\[2ex]
$^a$ {\it Theory Group, KEK, Tsukuba, Ibaraki 305-0801, Japan}    \\
$^b$ {\it Scuola Internazionale Superiore di Studi Avanzati 
      (SISSA), Via Beirut 4, I-34014 Trieste, Italy}              \\
$^c$ {\it Department of Physics, University of Tokyo,
          Tokyo 113-0033, Japan}                                  \\
$^d$ {\it Department of Physics, University of California, Berkelely, 
          California 94720, USA}                                  \\
$^e$ {\it Theoretical Physics Group, Lawrence Berkeley National 
          Laboratory, Berkelely, California 94720, USA}           \\
$^f$ {\it Research Center for the Early Universe,
          University of Tokyo, Tokyo 113-0033, Japan}             \\[8ex]
\end{center}
{\begin{center} ABSTRACT \end{center}}  
\vspace{-4truemm}
{\small{
\noindent 
The problem of generating light neutrinos within 
supersymmetric models is discussed. It is shown that 
the hierarchy of scales induced by supersymmetry breaking 
can give rise to suppression factors of the correct 
order of magnitude to produce experimentally allowed
neutrino spectra. 
}}

\section{Introduction}
\label{intro} 
Small neutrino masses are usually generated through the seesaw
mechanism~\cite{SEESAW}. Three superfields
$\bar{N}_\alpha$ ($\alpha = e,\mu,\tau$), singlets with respect to the
Standard Model (SM), are added to the three neutral state components
of the three doublets $L_\alpha$. These fields have a large mass $M_R$
that suppresses the electroweak scale naturally obtained from the
Yukawa operators $y_{\bar{N}} \bar{N} L H$, for coupling constants 
$y_{\bar{N}} = {\cal O}(1)$, when the neutral component of $H$
acquires a vacuum expectation value ({\it vev}).  Three light neutrino
eigenstates $\nu_1$, $\nu_2$, $\nu_3$, with mass $\sim v^2/M_R$ are
obtained, with very small mixing to the heavy eigenstates $n_1$,
$n_2$, $n_3$, with mass $\sim M_R$.

This mechanism can be easily implemented in supersymmetric models. In
these, however, given the richness with which they are endowed, it is
possible to obtain similar suppression factors in different ways, when
three SM singlet superfields $\bar{N}$ are added to the SM particle
content. In particular, supersymmetry breaking at some intermediate
scale $M_X$, with the typical hierarchy between the Planck mass $M_P$
($\simeq 10^{18}\,$GeV) and the gravitino mass $m_{3/2}$, 
$m_{3/2}\sim M_X^2/M_P$, or generically between the Planck mass and a
typical soft supersymmetry-breaking mass $\tilde{m}$ 
($m_{\rm weak} \ltap \tilde{m} \ll M_P $, where $m_{\rm weak}$ is the 
electroweak scale), may play a crucial role in providing alternative
mechanisms for generating small neutrino
masses~\cite{Borzumati:2000mc,Arkani-Hamed:2000bq,Arkani-Hamed:2000kj}.
Moreover, even keeping the seesaw mechanism unaltered, supersymmetry
breaking may still be advocated to explain the lightness of an
additional light sterile neutrino, if a fourth SM singlet superfield
$S$ is also included~\cite{Babu:2000hb}. The resulting seventh
neutrino eigenstate is denoted by $\nu_s$.

Here, we critically review mechanisms and models recently proposed in
Refs.~\cite{Borzumati:2000mc} and~\cite{Babu:2000hb}.  Both proposals
rely on the same class of models of supersymmetry
breaking~\cite{Izawa:1996pk,Intriligator:1996pu}.  They are, however,
very different in spirit. The proposal of Ref.~\cite{Borzumati:2000mc}
replaces or encompasses the conventional seesaw mechanism,
accommodating three states $n_i$ of mainly sterile neutrinos that are
light or heavy depending on the details of the specific model. The
proposal of Ref.~\cite{Babu:2000hb} makes use of the hierarchy of
scales induced by supersymmetry breaking to explain the lightness of
the fourth sterile neutrino.  The three states $\nu_i$, mainly active
ones, are made light by the well-known seesaw suppression generated in
the neutrino mass matrix by the three heavy sterile neutrinos of mass
$\sim M_R$.

When looking for alternatives to the seesaw mechanism, or when trying
to generate a small mass for a fourth sterile neutrino, an
excessive tuning of coupling constants can be avoided by forbidding
renormalizable interactions such as $y\bar{N} L H$ or $y_S SLH$.
Thus, neutrino masses can be generated (in addition to or as a
replacement of the seesaw mechanism) in the two following 
ways.

\begin{itemize}
\item 
{\bf Through non-renormalizable operators}, in which $\bar{N}$ and $S$ 
are coupled to a spurion field. Small masses are dynamically generated 
and linked to the large hierarchy between $m_{\rm weak}$ and $M_P$, or
between $m_{\rm weak}$ and some intermediate scale, as in the seesaw
mechanism. Contributions to Dirac masses $m_D$ and $m_{Ds}$ are
induced by non-renormalizable interactions of the Yukawa type,
i.e. terms of the form
$\bar{\nu}_{L\,\alpha} \nu_{R\,\beta}$, 
$\bar{\nu}_{R\,\alpha} \nu_{L\,\beta}$, 
and 
$\bar{\nu}_{L s} \nu_{L\,\alpha}$, 
$\bar{\nu}_{L\,\alpha} \nu_{L s}$, 
where
$\nu_{L\,\alpha}$ indicates the fermionic component of the neutral
fields in $L_{\alpha}$ (the $\alpha$-th active neutrino), $\nu_{L s}$
and $\nu_{R\,\alpha}$ are the charge conjugated fermionic components
of the superfields $S$ and $\bar{N}_\alpha$, respectively.  Majorana
masses $m_L$ for active neutrinos 
$\nu_{L\,\alpha}^T C \nu_{L\,\beta}$ can be generated, if 
chiral-lepton violation is allowed.

\item
{\bf Through quantum effects}. As in the previous mechanism, loop
diagrams give rise to Majorana masses for active neutrinos, if
chiral-lepton violation is allowed. Radiative contributions to Dirac
masses are possible if an additional gauge group, such as $U(1)_{B-L}$
is included.
\end{itemize}

In the following, we discuss the contribution from non-renormalizable 
operators to Dirac and Majorana entries in the neutrino mass matrix, 
see Sec.~\ref{non-renorm}.  The superpotentials for the relevant fields 
in the different models proposed in Refs.~\cite{Borzumati:2000mc}   
and~\cite{Babu:2000hb} are listed in 
Sects~\ref{bnmod_nolnv},~\ref{bnmod_explnv},~\ref{bnmodels_spontlnv}, 
and~\ref{by_model}. In Sec.~\ref{susy-break-models}, a class of models 
is outlined in which supersymmetry breaking is induced by the strong 
dynamics of a gauge group $SU(2)$, ${\cal F}_Z \ne 0$, where 
${\cal F}_Z$ is the {\it vev} of the auxiliary component of a singlet 
superfield $Z$. The breaking of supersymmetry gives rise to a 
non-vanishing {\it vev} for the scalar component of such a superfield,
${\cal A}_Z \ne 0$, which is crucial to induce Dirac and 
Majorana mass terms from the non-renormalizable operators listed in 
Sec.~\ref{non-renorm}. The same operators give rise to scalar 
terms softly breaking  supersymmetry, with explicit chiral-lepton 
violations.  These are discussed in Sec.~\ref{scalars}. Some of these
terms, if not duly suppressed (or altogether forbidden) may give rise 
to instability problems and/or too large radiative 
contributions to Majorana neutrino masses. At danger is 
the model of Ref.~\cite{Babu:2000hb}, for which remedies are
anticipated already in Sec.~\ref{by_model} and outlined in 
Secs.~\ref{stability} and~\ref{cosmology}.  Finally the 
contributions to Dirac and Majorana masses due to quantum effects, 
also induced by supersymmetry breaking, are discussed in 
Sec.~\ref{radiative-masses}. The typical neutrino spectra obtained 
in the models proposed in Refs.~\cite{Borzumati:2000mc} 
and~\cite{Babu:2000hb} are described in Sec.~\ref{spectra}.

\section{Contributions from non-renormalizable operators}
\label{non-renorm} 
We list some non-renormalizable operators inducing tree-level neutrino
masses. The presentation is arranged in such a way to show the basic
structure of the different models proposed in
Refs.~\cite{Borzumati:2000mc} and~\cite{Babu:2000hb}. Their
classification is according to the type of lepton-number violation in
them allowed.

\subsection{No gauge group extension; 3 singlets $\bar{N}$}
\label{bnmodels}
\subsubsection{No lepton-number violation allowed}
\label{bnmod_nolnv}
The relevant superpotential operator is:
\begin{equation}
 W =  \frac{k_{\bar{N}}}{M_P} Z \bar{N}L H \,,
\label{spot_one}
\end{equation}
where $Z$ is a SM singlet, with a
supersymmetry-conserving vacuum expectation value ({\it vev})
${\cal A}_Z$ and a supersymmetry-violating one ${\cal F}_Z$, with
${\cal F}_Z \ltap M_X^2 \simeq m_{3/2} M_P$. The usual Yukawa
coupling operator $y \bar{N} L H$ can be forbidden by a 
continuous or discrete symmetry $Z_n$:
\begin{equation}
 Z_n(\bar{N})=+1, \quad Z_n(Z) = -1    \,,
\label{znsymm}
\end{equation} 
whereas a Majorana mass for the right-handed neutrinos is forbidden by
a discrete lepton-number symmetry $L$:
\begin{equation}
  L(+1), \quad
  \bar{N}(-1), \quad
  \bar{E}(-1) \,.
\label{lepton-num}
\end{equation} 
For each generation, the Dirac neutrino mass 
\begin{equation}
 m_D = k_{\bar{N}} \,\frac{{\cal A}_Z v}{M_P}
\label{spectrum_one}
\end{equation} 
is generated and one Dirac neutrino eigenstate with mass 
$ m_\nu = m_D$ is induced. Notice that, if 
${\cal A}_Z^2  \simeq M_X^2 \simeq  m_{3/2} M_P$, 
the neutrino mass is too large, i.e. $m_\nu \simeq 10^3\,$eV. 
Thus, a value ${\cal A}_Z^2 \ll M_X^2 \ltap m_{3/2} M_P$
is needed.

\subsubsection{Explicit lepton-number violation}
\label{bnmod_explnv}
In this case, the superpotential is:
\begin{equation}
 W =  \frac{k_{\bar{N}}}{M_P}     Z  \bar{N}L H      \ + \
      \frac{1}{4} \frac{z}{M_P} Z Z \bar{N} \bar{N}  \ + \ 
      \frac{1}{4} \frac{h}{M_P} L H L H \,. 
\label{spot_two}
\end{equation}
The same $Z_n$-symmetry of eq.~(\ref{znsymm}) forbids the usual Yukawa
operators and the absence of a large Majorana mass for the fields
$\bar{N}$, can be guaranteed by imposing that $Z_n$ is $Z_{n\ne 2}$.
An additional $R$-parity symmetry, $R_p$:
\begin{equation}
 R_p(\bar{N})=-\,, \quad R_p(Z) = +    \,,
\label{rpsymm}
\end{equation} 
forbids also dangerous terms such as $M_P Z \bar{N}$. 
For each generation, the tree-level contribution to the Dirac mass
$m_D$, and the Majorana masses for left-handed and right-handed
neutrinos, $m_L$ and $M_R$ are, respectively:
\begin{equation}
 m_D = k_{\bar{N}} \,\frac{{\cal A}_Z v}{M_P}\,, 
\hspace*{0.5truecm}
 M_R =\frac{z}{2} \,\frac{{\cal A}_Z^2}{M_P} \,, 
\hspace*{0.5truecm}
 m_L =\frac{h}{2} \,\frac{v^2}{M_P} \,.
\label{spectrum_two}
\end{equation}
The physical neutrino spectrum consists of two Majorana neutrinos 
with mass:
\begin{equation}
 m_{\nu_1} \simeq \left(\!\frac{h}{2}\! -\!
                          \frac{2 k_{\bar{N}}^2}{z} \right)
                  \displaystyle{\frac{v^2}{M_P}}            \,,   
 \quad  \quad 
 m_{\nu_2} \simeq \frac{z}{2} \displaystyle{\frac{{\cal A}_Z^2}{M_P}} \,.
\end{equation}
The only assumption so far is that $v<{\cal A}_Z$ and that $z$ is
unsuppressed. In this case, for a value of 
${\cal A}_Z^2 \simeq M_X^2 \simeq m_{3/2} M_P$, $m_{\nu_1}$ is 
too small, i.e. $m_{\nu_1} \simeq 10^{-5}\,$eV, whereas 
$m_{\nu_2} \simeq m_{3/2}$. This spectrum may be cured by radiative
corrections, which can increase the value of $m_L$.

\subsection{Spontaneous lepton-number violation; 3 singlets $\bar{N}$}    
\label{bnmodels_spontlnv}
Spontaneous lepton-number violation is achieved through 
the spontaneous breaking of an additional gauge group, say a 
$U(1)_{B-L}$. Due to this new gauge interaction, two additional 
Higgs bosons, $\Phi$ and $\bar{\Phi}$ (SM singlets) are present. 
They acquire {\it vev}'s
$<\!\! \Phi\!> = <\! \bar{\Phi}\!> = v_\Phi = v_{\bar{\Phi}}$. 
The fields $\bar{N}$ are not neutral with respect to $U(1)_{B-L}$
and have charge $X_{\bar{N}}$, while the charges of $\Phi$ and 
$\bar{\Phi}$ are respectively $X_\Phi$ and $X_{\bar{\Phi}}$.   
As in the two cases described before, renormalizable neutrino Yukawa 
interactions are forbidden by the $Z_n$-symmetry 
in eq.~(\ref{znsymm}), and the relevant superpotential is:
\begin{equation}
 W =  \frac{k_{\bar{N}}}{M_P}         Z  \bar{N}L H  
\ + \
      \frac{1}{4} \frac{z}{(M_P)^{m+1}} \bar{\Phi}^m Z Z \bar{N} \bar{N}
\ + \ 
      \frac{1}{4} \frac{h}{(M_P)^{m+1}} {\Phi}^m  L H L H \,,
\label{spot_three}
\end{equation}
if the symmetry that forbids the Yukawa neutrino operator is 
$Z_{n\ne 2}$. The operator giving rise to the Majorana mass 
for the fields $\bar{N}$ is however: 
\begin{equation}
 W =  \frac{1}{2} \frac{z}{(M_P)^{m-1}} \bar{\Phi}^m  \bar{N} \bar{N}\,,
\label{spot_four}
\end{equation}
if $Z_n =Z_2$. The power $m$ in eqs.~(\ref{spot_three}) 
and~(\ref{spot_four})
is a solution of the equation $2 X_{\bar{N}} + m X_{\bar{\Phi}} = 0$. 
In the first case, i.e. $Z_{n\ne 2}$, 
the tree-level contributions to neutrino masses are:
\begin{equation}
 m_D = k_{\bar{N}} \,\frac{{\cal A}_Z v}{M_P}\,, 
\hspace*{0.5truecm}
 M_R =\frac{z}{2} \,\left(\frac{v_{\bar{\Phi}}}{M_P}\right)^m
                    \frac{{\cal A}_Z^2}{M_P} \,, 
\hspace*{0.5truecm}
 m_L =\frac{h}{2} \,\left(\frac{v_{\Phi}}{M_P}\right)^m
                    \frac{v^2}{M_P} \,. 
\label{spectrum_three}
\end{equation}
If the discrete symmetry is $Z_2$, $M_R$ becomes:
\begin{equation}
 M_R = z \,v_{\bar{\Phi}}\,
              \left(\frac{v_{\bar{\Phi}}}{M_P}\right)^{m-1} \,, 
\label{spectrum_four}
\end{equation}
whereas $m_D$ and $m_L$ remain unchanged. 
The physical
spectrum obtained depends on the values of $m$, $v_{\bar{\Phi}}$ and
${\cal A}_Z$.  For $Z_n =Z_2$, and $m=1$, it is 
$M_R \sim v_{\bar{\Phi}}$. In general, however, 
since
$v_{\bar{\Phi}} \le M_P$, the factor $({v_{\bar{\Phi}}}/{M_P})^m $ 
is a genuine suppression factor and $M_R$ in  
eq.~(\ref{spectrum_four}) is $< v_{\bar{\Phi}}$. The same is true 
for $Z_{n\ne 2}$, i.e. $M_R$ in eq.~(\ref{spectrum_three}), if 
${\cal A}_Z^2 \ltap M_X^2 \simeq m_{3/2} M_P$.
In particular, for sufficiently large $m$, it is $M_R \ll v_{\Phi}$. 
Moreover, if $v_{\bar{\Phi}}$ is sufficiently close to $M_P$, 
the spectrum is similar to that obtained with
the superpotential in eq.~(\ref{spot_two}). If $v_{\bar{\Phi}}$ is,
for example, as in the seesaw case, $\simeq 10^{12}\,$GeV, and $m=1$,
then, for ${\cal A}_Z^2 \simeq M_X^2 \simeq m_{3/2} M_P$, $\nu_2$
would tend to be too heavy, $\sim 10^{5}\,$eV, and 
$m_{\nu_1}\sim 10\,$eV. For larger $m$, the suppression factor becomes
increasingly more effective, and the two eigenvalues tend to 
$\pm k_{\bar{N}} {{\cal A}_Z v}/{M_P}$. Again, a solution to the
problem may be obtained if ${\cal A}_Z^2 \ll M_X^2$ is possible.
Similar considerations hold in the case of $Z_2$.

\subsection{Lepton-number violation; 4 singlets:
  3 $\bar{N}$'s, 1 $S$}   
\label{by_model}
This case is slightly different from the previous ones, in that
renormalizable operators that give rise to the usual seesaw mechanism
are allowed, whereas it is only the renormalizable Yukawa operator for
the fields $S$ that is forbidden. A non-renormalizable operator of the 
Yukawa type, mediated by a SM singlet $Z$, is allowed. 
The superpotential is, in this case:
\begin{equation}
 W =   y \bar{N} L H                    \ + \
      \frac{1}{2}M_R \bar{N} \bar{N}    \ + \ 
      \frac{1}{4}\frac{h}{M_P} L H L H  \ + \ 
      \frac{k_S}{M_P} Z S L H           \,,
\label{spot_five}
\end{equation}
where, for simplicity, lepton number is assumed to be explicitly
broken. The modifications for the case of spontaneous lepton-number
violation, which requires an additional gauge group, are
obvious. Notice that all bilinear mass terms $S S $ and $\bar{N} S$
are understood to be forbidden as in string scenarios in which $S$ is
a moduli field. It is also understood that some string-related 
dynamical feature forbids the non-renormalizable superpotential 
operator $({1}/{M_P}) ZZ SS$, which could otherwise induce 
very dangerous scalar interaction terms. 
Finally, by assuming that the fields relevant 
for the generation of neutrino masses are charged under a $U(1)_R$ 
group, with the following $R$-charges:
\begin{equation}
  R(Z) =0\,,      \quad  
  R(H) =0\,,      \quad 
  R(L) =1\,,      \quad 
  R(\bar{N})=1\,, \quad 
  R(S) =1 \,, 
\label{byRcharges}
\end{equation}
a similarly dangerous operator in the K\"ahler potential, 
$(1/M_P^2) Z^\dagger Z S S$, is also forbidden.   
The consequences that both operators, $({1}/{M_P}) ZZ SS$ 
and $(1/M_P^2) Z^\dagger Z S S$, could have are discussed 
in Sec.~\ref{scalars}.

By integrating out the heavy fields $\bar{N}$, the usual effective
superpotential
\begin{equation}
 W = -\frac{y^2}{2 M_R} L H L H           \ + \
      \frac{k_S}{M_P} Z S L H           \,,
\label{spot_five_a}
\end{equation}
is obtained, where the subleading operator, with coupling $h$, is 
neglected. Four light states are now present, three with Majorana 
masses $m_L$, one with Dirac mass $m_{Ds}$:
\begin{equation}
 m_{Ds} = k_S \,\frac{{\cal A}_Z v}{M_P}\,,  
\hspace*{0.5truecm}
 m_L      = - y^2 \frac{v^2}{M_R} \,. 
\end{equation}
Values of $M_R \sim 10^{13}$--$10^{14}$ give rise to $m_L$'s in 
the range $0.1$--$1\,$eV, whereas $m_{Ds}$ is again too large if 
${\cal A}_Z^2 \simeq M_X^2 \simeq m_{3/2} M_P$.

\section{An interesting class of models}
\label{susy-break-models}
An interesting class of models was introduced in the past years, in
which the dynamics of a strongly interacting $SU(2)$ gauge group
induces the breaking of supersymmetry through non-perturbative
effects~\cite{Izawa:1996pk,Intriligator:1996pu}. The scale of
supersymmetry breaking coincides with the dynamical scale $\Lambda$ of
this gauge interaction. A fundamental role is played by a superfield
$Z$, singlet of $SU(2)$ and of the SM gauge group, which is the field
that acquires a supersymmetry-breaking {\it vev} 
${\cal F}_Z \sim \Lambda^2 $. (For the definition of the field $Z$,
see Ref.~\cite{Izawa:1997gs}.) This field is identified with the
singlet mediating the non-renormalizable neutrino operators in
eqs.~(\ref{spot_one}),~(\ref{spot_two}),~(\ref{spot_three}),
and~(\ref{spot_five}).

It is known that in these models the $Z$ direction is flat at the tree
level, but it is lifted by loop corrections.  Corrections to the
K\"ahler potential from the strong $SU(2)$ interaction are
non-calculable and can only be estimated. They give rise to a
quadratic term in $Z$ in the scalar potential, with coupling
$k$. Making a specific dynamical assumption on the sign of this term,
i.e. $k>0$, a tiny {\it vev} for $Z$ is 
induced~\cite{Izawa:1995jg}:
\begin{equation}
 {\cal A}_Z \simeq \frac{m_{3/2}}{\lambda k}  \,, 
\label{nonpertAz} 
\end{equation}
where $\lambda$ is the coupling of the effective low-energy 
superpotential obtained after integrating out heavy fields 
active under the strong $SU(2)$. It was shown 
in~\cite{Chacko:1998si}, however, that a quadratic term in the 
scalar potential is obtained from loops of light 
particles. These corrections are calculable and larger than the
non-calculable one as long as $\lambda$ is in a perturbative
regime~\cite{Chacko:1998si}. This quadratic term induces a 
{\it vev}:  
\begin{equation}
{\cal A}_Z \simeq \frac{16\pi^2}{\lambda^3} m_{3/2}  \,.
\label{pertAz}
\end{equation}
In the following, $\lambda$ is assumed to be in the perturbative 
regime and only the value in eq.~(\ref{pertAz}) is used for the 
{\it vev} ${\cal A}_Z$. (The numerical result obtained in
 Ref.~\cite{Babu:2000hb} for the mass of the sterile neutrino, makes
 use of the value of ${\cal A}_Z$ in eq.~(\ref{nonpertAz}),
 corresponding to a non-perturbative coupling $\lambda$.)

The ratio ${\cal A}_Z/M_P$ in
eqs.~(\ref{spectrum_one}),~(\ref{spectrum_two}),
and~(\ref{spectrum_three}) may be sufficiently small to induce
suitable values of neutrino masses. Indeed, when the breaking of
supersymmetry is transmitted to the SM visible sector via
gravitational interactions ($m_{3/2} \simeq 1\,$TeV), the ratio 
${\cal A}_Z/M_P$ is of order $\sim 10^{-13}$. It is needless to say
that, if the transmission of supersymmetry breaking is via gauge
interactions (in general, $m_{3/2} \ll m_{\rm weak}$), this ratio 
may give rise to contributions to neutrino masses that are too small. 
The radiative mechanism may then be the only one to generate
sizable values of $m_L$ and $m_D$ in this class of models.

Notice that in the case of the superpotentials in
eqs.~(\ref{spot_one}),~(\ref{spot_two}), and~(\ref{spot_three}), the 
singlet $Z$ is charged under the discrete group $Z_n$, while it is
neutral in the physics picture described by the superpotential in
eq.~(\ref{spot_five}). In this picture, $Z$ can couple to vector
multiplets, giving rise to gaugino masses and can therefore be the
main supersymmetry-breaking agent, when the information of
supersymmetry breaking is transmitted to the SM visible sector via
gravitational interactions.  It is assumed in Ref.~\cite{Babu:2000hb}
that ${\cal F}_Z \sim \Lambda^2 \sim M_X^2 \sim m_{3/2} M_P$.
In the scenarios described by the superpotential in
eqs.~(\ref{spot_one}),~(\ref{spot_two}), and~(\ref{spot_three}), $Z$
cannot induce gaugino masses and it does not need to be the singlet
with the largest supersymmetry-breaking {\it vev}.  Thus, it can be
${\cal F}_Z \sim \Lambda^2 < M_X^2 \sim m_{3/2} M_P$.  The situation
is obviously more free when the breaking of supersymmetry is
transmitted to the visible sector by gauge interactions, which induce
gaugino masses at the quantum level. In this case, ${\cal F}_Z$ can 
be smaller (and even considerably smaller) than $M_X^2$.

\section{Scalar interactions} 
\label{scalars}
In addition to the mass terms 
$\tilde{m}_{\bar{N}}^2   \vert \tilde{\bar{N}}\vert^2$ and 
$\tilde{m}_S^2 \vert \tilde{S} \vert^2 $ induced by 
K\"ahler potential operators, trilinear as well as bilinear 
terms holomorphic in the scalar components of $\bar{N}$ and $S$ 
are also generated.

The non-renormalizable operators $(1/M_P) Z \bar{N} L H$ and 
$(1/M_P) Z S L H$ give rise to the terms:
\begin{equation}
 A_{\bar{N}} \tilde{\bar{N}} \tilde{L} H \,, \quad \quad 
 A_S \tilde{S} \tilde{L} H       \,, 
\label{trilinear}
\end{equation}
with dimensionful couplings given by:
\begin{equation}
 A_{\bar{N}} = k_{\bar{N}} 
                   \left(\frac{{\cal F}_Z}{M_P}\right)\,, \quad \quad 
 A_S         = k_S \left(\frac{{\cal F}_Z}{M_P}\right)\,.
\end{equation}
These are of order of the gravitino mass if ${\cal F}_Z$ is the
maximal supersymmetry-breaking {\it vev}, and therefore 
$\gtap m_{\rm weak}$ in gravity-mediated scenarios 
of transmission of supersymmetry breaking, for couplings $k_{\bar{N}}$ 
and $k_S$ of ${\cal O}(1)$. $A_{\bar{N}}$ can be much smaller than 
$m_{\rm weak}$ when this assumption is dropped, i.e. when 
${\cal F}_Z < M_X^2$, or in gauge-mediated scenarios of 
supersymmetry breaking.

Bilinear terms, such as: 
\begin{equation}
 B_{\bar{N}}^2 \tilde{\bar{N}} \tilde{\bar{N}}\,, \quad \quad 
 B_S^2         \tilde{S}       \tilde{S}      \,, 
\label{bilinear}
\end{equation}
are also generated, the first in the models of
Secs.~\ref{bnmod_explnv}, and~\ref{bnmodels_spontlnv} (no such term 
is possible in the model of Sec.~\ref{bnmod_nolnv}, in which lepton
number is conserved), the second in the model described in
Sec.~\ref{by_model}.

In the models of Secs.~\ref{bnmod_explnv},
and~\ref{bnmodels_spontlnv}, in which lepton number is violated
explicitly or spontaneously, due to the presence of additional
symmetries, a supersymmetric mass term $(1/2) M_R N N $ is induced by
non-renormalizable operators, Thus, a bilinear term
\begin{equation}
 B_{\bar{N}}^2 = m_{3/2} M_R
\label{convbterms}
\end{equation} 
is always generated via supergravity effects. Moreover, in the model
of Secs.~\ref{bnmod_explnv}, and that of Sec.~\ref{bnmodels_spontlnv}
with $Z_{n\ne 2}$, a term
\begin{equation}
 B_{\bar{N}}^2 = \frac{{\cal F}_Z}{{\cal A}_Z} M_R
\label{additterms}
\end{equation}
is directly induced by the relevant non-renormalizable terms in the
superpotentials of eqs.~(\ref{spot_two}) and~(\ref{spot_three}). In 
gravity-mediated scenarios of transmission of supersymmetry breaking,
no instabilities or tachyonic eigenvalues for the scalar components 
of $\bar{N}$ arise, as far as $B_{\bar{N}}^2 < M_R^2$, if 
$M_R \gtap \tilde{m}_{\tilde{N}}$, or as far as 
$B_{\bar{N}}^2 < \tilde{m}_{\tilde{N}}^2$, if 
$M_R < \tilde{m}_{\tilde{N}}$.  All potentially dangerous situations 
can be avoided by requiring ${\cal F}_Z < M_X^2 \sim m_{3/2} M_P$. 
(See discussion in Ref.~\cite{Borzumati:2000mc} and in the previous 
Section.) No problems in general arise in scenarios in which the
transmission of supersymmetry breaking is via gauge interactions.

In the model of Sec.~\ref{by_model}, chiral-lepton number is violated
through the field $\bar{N}$, but a renormalizable mass term for the
field $S$ is forbidden by ``stringy'' reasons. However, if no charges
are assigned to any of the relevant fields, it is difficult to forbid
the non-renormalizable operator $(1/M_P) ZZ SS$ already mentioned in
Sec.~\ref{by_model}.  Indeed, the combination of fields $Z S $ has, in
principle, the same quantum numbers of the field $\bar{N}$, unless
some higher scale string dynamics is advocated to distinguish between
$ZS $ and $\bar{N}$. Thus, forbidding $(1/M_P) ZZ SS$ would be
equivalent to forbidding the see-saw operator $(1/M_R) L H L H $. Such
an operator would give rise to an acceptable tree-level Majorana mass
for the sterile neutrino, of order
${\cal A}_Z^2/M_P \sim (16 \pi^2/\lambda^3)^2 m_{3/2}^2 /M_P$, 
with mixing $(v/m_{3/2})$ to the active neutrinos.  It would, however,
also give rise to a tachyonic scalar component of the field $S$, and
to a very large radiative contributions to the Majorana mass of active
neutrinos, through the large bilinear term
\begin{equation}
 B_S^2   = \frac{16 \pi^2}{\lambda^3} m_{3/2}^2 \,.
\label{BSfromW}
\end{equation}
(See discussion in Sec.~\ref{radiative-masses}.) Similarly, the
K\"ahler potential operator $(1/M_P^2) Z^\dagger ZSS$, would induce
the dangerous value of $B_S^2$:
\begin{equation}
 B_S^2   = m_{3/2}^2  \,.  
\label{BSfromK}
\end{equation}
The simple assignment of $R$-charges in eq.~(\ref{byRcharges}) can,
however, forbid this operator.

\section{Stability of the scalar potential}
\label{stability}
We start by discussing the scalar potential relevant for the model 
described in Sec.~\ref{by_model}. It is easy to see that this 
potential has a $D$-flat direction when:
\begin{equation}
 \tilde{L}= \frac{1}{\sqrt{2}} \left(\! 
    \begin{array}{c} 0 \\ \phi \end{array} \! \right) \,, \quad 
     H    = \frac{1}{\sqrt{2}} \left(\! 
    \begin{array}{c} \phi \\ 0 \end{array} \! \right) \,, 
\end{equation}
where $H$ is here understood as the scalar component of the superfield
indicated by the same symbol. For a choice of phases such that 
$\phi^2 \tilde{S} = -\vert \phi \vert^2 \vert \tilde{S} \vert$, the
scalar potential for $\tilde{S}$ and $\phi$ around the origin, is
given by:
\begin{eqnarray}
 V = \tilde{m}_{\phi}^2 |\phi|^2 + \tilde{m}_S^2 |\tilde{S}|^2 - 
      2 k_S \, m_{3/2} \vert \phi \vert^2 \vert \tilde{S}\vert 
      + \cdots                                                   \,.
\label{eqV}
\end{eqnarray}
In this expression, the ellipsis denotes higher order terms, 
suppressed by small factors of $(m_{\rm weak}/M_R)$, 
$(m_{\rm weak}/M_P)$, or $({\cal A}_Z /M_P)$, as long as the field 
values $|\phi|$ and $|\tilde{S}|$ are not too large compared to the
electroweak scale. It is clear that such a potential has a minimum
deeper than the origin for values of the field $|\phi|$ and
$|\tilde{S}|$ larger than the electroweak scale. In fact, the
potential in eq.~(\ref{eqV}) crosses the plane $V = 0$ at the
following field values:
\begin{eqnarray}
 |\tilde{S}_0| &\simeq& \frac{1}{2 k_S}
  \left(\frac{\tilde{m}_{\phi}^2}{m_{3/2}}\right)(1+a) \,,
  \\
 |\phi_0| &\simeq& \frac{1}{2 k_S}
  \left( \frac{\tilde{m}_{\phi} \tilde{m}_S}{m_{3/2}} \right)
  \left(\frac{1+a}{\sqrt{a}}\right)   \,,
\end{eqnarray}
where $a$ is an arbitrary real number. These values are of the order
of weak scale and tunneling to the unwanted minimum may occur.

However, an estimate of the tunneling rate to the false vacuum shows
that a value of the coupling $k_S< 1$ by about one order of
magnitude, is sufficient to guarantee that the origin is a metastable
vacuum.  The tunnelling rate can be estimated from the four
dimensional Euclidean action $S_4$ evaluated with the bounce solution
for the potential in eq.~(\ref{eqV})~\cite{COLEMAN}. (See also 
discussion in~\cite{Kawasaki:2000ye}.) The tunneling rate per volume 
is roughly given by 
$\Gamma_4 \sim {\tilde m}^4\exp(-S_4)$, where $\tilde{m}$ is a 
typical scale of the potential,
$\tilde{m} \sim \tilde{m}_{\phi}\sim \tilde{m}_S \sim m_{3/2}$. The 
requirement that the false vacuum has not decayed in our past
light-cone gives a constraint $\Gamma_4 L^4 \ll 1$, where $L$ is the
present size and age of the visible universe. Thus, $S_4$ must satisfy
$S_4 > 400 + 4\ln\left( \tilde{m} / {\rm TeV}\right)$.  On the other
hand, from eq.~(\ref{eqV}), by redefining $\phi$ and $\tilde{S}$, one
can find $S_4 \sim (1/k_S^2) \times \hat{S}_4$, where $\hat{S}_4$ is a
dimensionless numerical factor of 
${\cal O}(10)$~\cite{Linde:1983zj}. Therefore, a value $k_S \ltap 0.1$ 
is sufficient to avoid tunneling to the dangerous vacuum.
Even if $k_s \simeq 0.01$, we may reproduce the result in
Ref.~\cite{Babu:2000hb} adopting the value ${\cal A}_Z$ in
eq.~(\ref{pertAz}).

Notice that a similar problem may arise in the models described in
Secs.~\ref{bnmodels} and~\ref{bnmodels_spontlnv}, except for the 
case in which an unsuppressed mass term for the superfield $N$ is 
present (see the superpotential in eq.~(\ref{spot_four}) with 
$m=1$). With a choice of phases for $\phi$ and $\tilde{\bar{N}}$ 
similar to that made earlier for $\phi$ and $\tilde{S}$, the 
scalar potential is in these cases:
\begin{eqnarray}
 V = \tilde{m}_{\phi}^2 |\phi|^2 +
     \tilde{m}_{\bar{N}}^2 \vert\tilde{\bar{N}}\vert^2 - 
      2 k_{\bar{N}} \left(\frac{{\cal F}_Z}{M_P}\right)
      \vert \phi \vert^2 \vert \tilde{\bar{N}}\vert 
      + \cdots                                                   \,.
\label{eqV_BN}
\end{eqnarray}
As mentioned earlier, however, in these models the singlet $Z$ does
not need to have the maximal value of supersymmetry-breaking 
${\it vev}$ ${\cal F}_Z$, as it is assumed in the model of 
Sec.~\ref{by_model}. Therefore, a suppression of the trilinear term 
$\vert \phi\vert^2 \vert \tilde{\bar{N}}\vert$ may be achieved 
through a smaller value of ${\cal F}_Z$, while keeping the 
coefficient $k_{\bar{N}} \sim 1$. (This and other types
 of solutions were advocated for the analogous problem present in 
 scenarios in which tree-level Yukawa couplings for quarks and 
 leptons are forbidden by symmetries~\cite{BFPT}.)

\section{Another cosmological problem}
\label{cosmology}
It was shown in the previous Section that the scalar potential in
eq.~(\ref{eqV}), as well as that in eq.~(\ref{eqV_BN}), if 
${\cal F}_Z/M_P \sim m_{3/2}$, have minima deeper than the origin.
The vacuum tunneling from the origin to these minima can be easily
avoided. However, another cosmological problem may, in principle,
arise.  If the values of the scalar fields $\tilde{S}$ or
$\tilde{\bar{N}}$ are larger than the weak scale at the end of
inflation, they do not roll toward the vacuum in which we live, but
towards the unwanted minimum, which is far from the origin, and the
desired vacuum is never realized.  However, if a scalar field is
charged, it is naturally expected that at the end of inflation such a
field is localized at the origin, which is an enhanced symmetry
point~\cite{WHO-2}.  The fields $\tilde{\bar{N}}$ are charged in all
the models outlined above (see Ref.~\cite{Borzumati:2000mc}) and this
problem is naturally avoided. The $R$-charges introduced in
eq.~(\ref{byRcharges}) to forbid a dangerous K\"ahler potential
operator, turns out to be very important to prevent this possible
cosmological problem, by forcing the field $S$ at the origin.

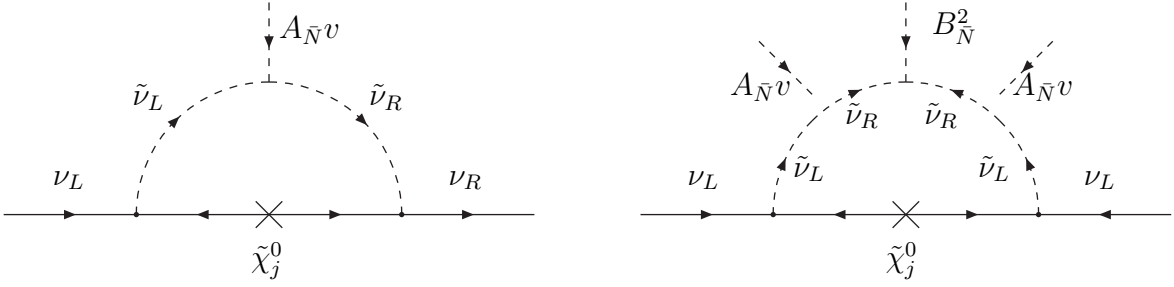
\begin{figure}[t]
\begin{center} 
\vspace*{0.5truecm}
\begin{picture}(100,80)(250,150)
  \ArrowLine(80,150)(130,150) \Text(105,160)[b]{$\nu_L$}
  \ArrowLine(180,150)(130,150)
  \ArrowLine(180,150)(230,150)
  \Line(175,145)(185,155) \Line(175,155)(185,145) 
  \Text(180,140)[t]{$\tilde{\chi}_j^0$} 
  \DashArrowArcn(180,150)(50,90,0){3}   \Text(135,195)[]{$\tilde{\nu}_L$}
  \DashArrowArcn(180,150)(50,180,90){3} \Text(225,195)[]{$\tilde{\nu}_R$}
  \DashArrowLine(180,230)(180,200){3}   \Text(196,215)[b]{$A_{\bar{N}} v$}
  \ArrowLine(230,150)(280,150) \Text(255,160)[b]{$\nu_R$}
  \Vertex(130,150){1}
  \Vertex(230,150){1}
 \ArrowLine(320,150)(370,150) \Text(345,160)[b]{$\nu_L$}
 \ArrowLine(420,150)(370,150)
 \ArrowLine(420,150)(470,150)
 \Line(415,145)(425,155) \Line(415,155)(425,145) 
 \Text(420,140)[t]{$\tilde{\chi}_j^0$} 
 \DashArrowArc(420,150)(50,0,45){3}    
               \Text(385,168)[]{$\tilde{\nu}_L$}
 \DashArrowArc(420,150)(50,45,90){3}   
               \Text(405,188)[]{$\tilde{\nu}_R$}
 \DashArrowArcn(420,150)(50,180,135){3}
               \Text(435,188)[]{$\tilde{\nu}_R$}
 \DashArrowArcn(420,150)(50,135,90){3} 
               \Text(455,168)[]{$\tilde{\nu}_L$}
 \DashArrowLine(420,230)(420,200){3} \Text(440,215)[b]{$B_{\bar{N}}^2 $}
 \DashArrowLine(364.64,215.36)(384.64,195.36){3} 
               \Text(366.64,193.36)[b]{$A_{\bar{N}} v$}
 \DashArrowLine(475.36,215.36)(455.36,195.36){3} 
               \Text(473.36,193.36)[b]{$A_{\bar{N}} v$}
 \ArrowLine(520,150)(470,150) \Text(495,160)[b]{$\nu_L$}
 \Vertex(370,150){1}
 \Vertex(470,150){1}
\end{picture}
\vspace*{0.5truecm}
\caption{Diagrams contributing to the Dirac and Majorana 
 neutrino masses $m_D$ and $m_L$.}
\label{diagsN}
\end{center}
\end{figure}

\section{Radiative contributions to neutrino masses}
\label{radiative-masses}
\begin{figure}[t]
\begin{center} 
\vspace*{0.3truecm}
\begin{picture}(100,80)(150,130)
 \ArrowLine(100,150)(150,150) \Text(125,160)[b]{$\nu_L$}
 \ArrowLine(200,150)(150,150) 
 \ArrowLine(200,150)(250,150) 
 \Line(195,145)(205,155) \Line(195,155)(205,145) 
 \Text(200,140)[t]{$\tilde{\chi}_j^0$} 
 \DashArrowArc(200,150)(50,0,45){3}    
               \Text(165,168)[]{$\tilde{\nu}_L$}
 \DashArrowArc(200,150)(50,45,90){3}   
               \Text(185,188)[]{$\tilde{\nu}_{L s}$}
 \DashArrowArcn(200,150)(50,180,135){3}
               \Text(215,188)[]{$\tilde{\nu}_{L s}$}
 \DashArrowArcn(200,150)(50,135,90){3} 
               \Text(235,168)[]{$\tilde{\nu}_L$}
 \DashArrowLine(200,230)(200,200){3} \Text(215,215)[b]{$B_S^2$}
 \DashArrowLine(144.64,205.36)(164.64,185.36){3} 
               \Text(146.64,183.36)[b]{$A_S v$}
 \DashArrowLine(255.36,205.36)(235.36,185.36){3} 
               \Text(253.36,183.36)[b]{$A_S v$}
 \ArrowLine(300,150)(250,150) \Text(275,160)[b]{$\nu_L$}
 \Vertex(150,150){1} 
 \Vertex(250,150){1} 
\end{picture}
\vspace*{0.3truecm}
\caption{Radiative contribution to $m_L$ from the sterile 
neutrino field $S$.}
\label{diagS}
\end{center}
\end{figure}
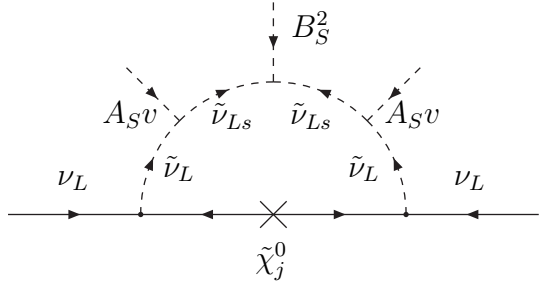

Radiative contributions to Dirac masses $m_D$ and Majorana masses
$m_L$ arise from the two diagrams shown in Fig.~\ref{diagsN} and/or
the diagram in Fig.~\ref{diagS}, depending on the model discussed. The
first diagram in Fig.~\ref{diagsN} requires a gauge interaction for
the field $\bar{N}$ and therefore is present only when the SM gauge
group is extended (see Sec.~\ref{bnmodels_spontlnv}).  The two
diagrams giving rise to the Majorana masses $m_L$ are possible when
chiral-lepton number is violated, specifically if bilinear terms
$B_{\bar{N}}^2 \tilde{\bar{N}} \tilde{\bar{N}}$ and 
$B_S^2 \tilde{S} \tilde{S}$ exist.

An explicit evaluation of these loop diagrams yields the following
results. The radiative contribution to Dirac masses $m_D$ is:
\begin{equation}
 m_D  \sim \frac{1}{8\pi^2} 
  \left(g_X X_{\bar{N}}\right) \left(g_Y Y_{L}\right)
  \left(\frac{A_{\bar{N}} v }{\tilde{m}} \right) 
  \left(\frac{\tilde{m}}{M_G}\right)^2 
\label{radDirac}
\end{equation}
where $g_X$ and $g_Y$ are respectively the coupling constants 
of the additional gauge group $U(1)_{B-L}$ and of $U(1)_Y$; 
$X_{\bar{N}}$ is the $U(1)_{B-L}$ charge of the superfields 
$\bar{N}$, $Y_L$ is the $U(1)_Y$ charge of the doublet superfield
$L$. Finally, $M_G$ is the scale of $U(1)_{B-L}$ violation:
$M_G \propto v_{\Phi} = v_{\bar{\Phi}}$ and $\tilde{m}$ is the 
typical soft mass for scalar fields and light gauginos.

The radiative contribution to Majorana masses $m_L$
shown in the second diagram of Fig.~\ref{diagsN}, i.e. 
due to the exchange of $\bar{N}$, is:
\begin{equation}
 m_L  \sim \frac{1}{8\pi^2} 
  \left(g_Y Y_{L}\right)^2
  \left(\frac{A_{\bar{N}} v }{\tilde{m}^2} \right)^2 
  \left(\frac{B M_R}{\tilde{m}}\right) 
  \left(\frac{\tilde{m}}{M_R}\right)^4  
\label{radMaj_largeN}
\end{equation}
if the Majorana mass for the fields $\bar{N}$ is 
$M_R \gg m_{\rm weak}$, whereas it is:
\begin{equation}
 m_L  \sim \frac{1}{8\pi^2} 
  \left(g_Y Y_{L}\right)^2
  \left(\frac{A_{\bar{N}} v }{\tilde{m}^2} \right)^2 
  \left(\frac{B M_R}{\tilde{m}}\right) \,,
\label{radMaj_smallN}
\end{equation}
for $M_R \ll m_{\rm weak}$. Notice that the symbol 
$B_{\bar{N}}^2$ in the Figure is here replaced by $B M_R$. 

It is interesting to see that in a seesaw type of setting, i.e. when
$M_G$ and $M_R$ are very large compared to $\tilde{m}$, $m_L$ and
$m_D$, and therefore the final mass eigenvalues, are more suppressed
by the large scales than in the typical seesaw mechanism. A part for a
loop suppression factor $(1/8\pi^2)$, they are reduced away from the
electroweak scale respectively by factors $(\tilde{m}/M_G)^2$ and
$(\tilde{m}/M_R)^3$. If the tree-level contributions to $m_D$ and
$m_L$ are forbidden as in the model of Sec.~\ref{bnmodels_spontlnv},
the scale of $U(1)_{B-L}$ can be much smaller than in the typical
seesaw mechanism.  Notice that in this model $M_G$ and $M_R$ are in
general disentangled and fixing the value of one ($M_G$) to get
reasonable values of $m_D$, does not necessarily imply that $m_L$ is
very small.

If $M_R$ is very small, and $B_{\bar{N}}^2 = B M_R$ is also small, 
the suppression to $m_L$ comes from $(B M_R/\tilde{m}^2)$ and 
$A_{\bar{N}}/\tilde{m}$. This is particularly true of the model 
of Sec.~\ref{bnmod_explnv} for which the options 
${\cal F}_Z < M_X^2$, or $m_{3/2} < m_{\rm weak}$ as when 
the transmission of supersymmetry breaking is mediated by gauge 
interactions, turns out to be of crucial importance.

\vspace*{0.5truecm}

In the model of Sec.~\ref{by_model}, since $M_R$ and possibly $M_G$
(if the model is realized in such a way to have spontaneous violation
of lepton number) are large, i.e.  $\sim 10^{13}\,$GeV, the radiative
contributions to $m_D$ and $m_L$ from the two diagrams of
Fig.~\ref{diagsN} are very small.  A radiative contribution to $m_L$,
however, arises from from the exchange of $\tilde{S}$ as shown in
Fig.~\ref{diagS}.  The natural scale for $\tilde{S}$ is the soft mass
$\tilde{m}$. Therefore, the expression for the radiative contribution
to $m_L$ is similar to that in eq.~(\ref{radMaj_smallN}):
\begin{equation}
 m_L  \sim \frac{1}{8\pi^2} 
  \left(g_Y Y_{L}\right)^2
  \left(\frac{A_{S} v }{\tilde{m}^2} \right)^2 
  \left(\frac{B_S^2}{\tilde{m}}\right) 
\label{radMaj_S}
\end{equation}
Since in this model ${\cal F}_Z$ is assumed to be the maximal
supersymmetry breaking ${\it vev}$, ${\cal F}_Z \sim M_X^2$, it is
$A_S \ltap \tilde{m}$ when the suppression of order of magnitude for
the coupling $k_S$ discussed in Sec.~\ref{stability} is kept into
account. It becomes clear, then, why the values of $B_S^2$ given in
eq.~(\ref{BSfromW}) and~(\ref{BSfromK}) are far too dangerous: indeed,
they tend to move the value of $m_L$ towards the electroweak scale ! 
It is therefore mandatory to have all operators involving the bilinear
combination $SS$ and giving rise to the soft term $B_S^2$ sufficiently
suppressed in order to avoid too large contributions to $m_L$.

\section{Neutrino spectra and Conclusions}
\label{spectra}
The neutrino spectrum obtained in the model of Sec.~\ref{by_model} is
that obtained from the conventional seesaw mechanism with the addition
of one light state, mainly sterile. There are, therefore, three heavy
states $n_i$ with mass $\sim M_R$, and four light states $\nu_i$ and
$\nu_s$.  (We assume here that radiative contributions to $m_L$ are
negligible.) In the basis in which charged leptons have a diagonal
mass matrix, for $k_S \sim 0.1$--$0.01$, as argued in
Sec.~\ref{stability}, and the value of ${\cal A}_Z$ in
eq.~(\ref{pertAz}), it is $m_{Ds} \sim 10^{-4}\,$eV. The values of 
$m_L$ are $\sim y^2 \times 1\,$eV. Therefore, if $\nu_e$ has a
Majorana mass $m_L \sim 10^{-4}\,$eV, the solar $\nu_e$-$\nu_{Ls}$ 
oscillation can be explained as a quasi-vacuum oscillation. With
suitable choices of the couplings $y$, the other two active neutrinos
together with $\nu_e$ can explain the atmospheric and LSND
oscillations.  The overall picture is that of the so-called $2\!+\!2$
neutrino spectrum~\cite{Kayser:2000ka}. (A model for the recently
 advocated ``$1\!+\!3$'' picture of neutrino
 masses~\cite{Barger:2000ch,Kayser:2000ka,Peres:2000ic}, in which all
 neutrino masses are also generated at the tree level, is proposed in
 Ref.~\cite{Borzumati:2000fe}.)

The model in Sec.~\ref{bnmod_nolnv} give rise to only three light
neutrinos with Dirac mass given in eq.~(\ref{spectrum_one}).  The
oscillation patterns of solar and atmospheric neutrino experiments
have to rely on a specific texture of the couplings $k_{\bar{N}}$ and
no explanation is possible for the LSND experiment.

The models of Secs.~\ref{bnmod_explnv} and~\ref{bnmodels_spontlnv} are
more complex and allow a large variety of neutrino spectra.  They have
an interesting interplay among tree-level and radiative contributions
to neutrino masses and replace completely the conventional seesaw
mechanism.

The model (or class of models) in Sec.~\ref{bnmodels_spontlnv} 
may nevertheless retain some of the qualitative feature of the 
seesaw mechanism spectrum if the Majorana mass $M_R$ for the fields 
$\bar{N}$ is large. Three light states $\nu_i$ are present, of 
mainly active neutrinos, and three heavy states $n_i$ with mass 
$\sim M_R$.  In this case, however, since no tree-level Yukawa 
couplings exist, the small mass of the 
light states $\nu_i$ is not due to the pattern of light-heavy scales 
in the neutrino mass matrix. What moves away from the 
electroweak scale ($m_{\rm weak} \sim v$) the Dirac and Majorana 
entries in the neutrino mass matrix are:
{\it i)} suppression factors coming from non-renormalizable operators,
i.e. ${\cal A}_Z/M_P$ and ${\cal A}_Z^2/(v M_P)$ where ${\cal A}_Z$ is
a vacuum expectation value induced by supersymmetry breaking;
{\it ii)} the factors $1/(16 \pi^2) (\tilde{m}/M_R)^3$ and $1/(16
\pi^2) (\tilde{m}/M_G)^2$, originated by loop diagrams, where
$\tilde{m}$ is a typical mass softly breaking supersymmetry and $M_G$
is the scale of the additional gauge group present in the model.

Both classes of models in Secs.~\ref{bnmod_explnv}
and~\ref{bnmodels_spontlnv}, however, may have rather small Majorana
masses $M_R$ at the tree-level and can easily realize the possibility
of light sterile neutrinos.  Exact predictions depend on the
particular realization of these models. The six states $\nu_i$ and
$n_i$ are all light.  The three $n_i$-neutrinos, however, can still be
heavier than the $\nu_i$'s, in which case, the mixing angles between
$n_i$'s and $\nu_i$'s are very small.  There is also the other
interesting case in which the $\nu_i$ and $n_i$ states are roughly at
the same scale or nearly degenerate, with mixing angles ranging from
small to maximal.  Flavour oscillations, therefore, can be
accommodated in this class of scenarios, relying on specific textures
of the tree-level couplings of non-renormalizable operators, or of the
trilinear and bilinear neutrino soft parameters.  Oscillations among
active and sterile neutrinos are also possible.

Since neutrino masses are much smaller than all other masses, it is
plausible to assume that their origin is different from that of the
other lepton and quark masses. They may be induced, after electroweak
symmetry breaking, by operators of higher dimensionality than the
usual Yukawa operators or by quantum effects.  We have shown that both
possibilities are easily implemented in supersymmetric models and that
the generation of neutrino masses is intimately linked to the breaking
of supersymmetry. In this sense, neutrinos may be fundamentally
different from all other fermions. The recently gathered evidence
pointing to the fact that they are massive and the pattern of
oscillation among different neutrinos, may therefore be the first
evidence for supersymmetry.

\vspace*{0.5truecm}
\noindent 
{\bf Acknowledgements}\\ 
The authors thank M.~Fujii and T.~Watari for discussions. 
F.B. is supported in part by a Japanese Monbusho fellowship.
She thanks the KEK theory group, in particular Y.~Okada, the Tokyo 
University theory group, and the CERN theory division for 
hospitality.  
The work of K.H. was supported by the Japanese Society for the 
Promotion of Science. 
Y.N. thanks the Miller Institute for Basic Research in Science 
for financial support.
T.Y. acknowledges partial support from the Grant-in-Aid for 
Scientific Research from the Ministry of Education, Sports, and 
Culture of Japan, on Priority Aerea \# 707: 
``Supersymmetry and Unified Theory of Elementary Particles''.

\end{document}